\documentstyle[psfig,twoside,fleqn,espcrc2]{article}

\title{Phase Transition of 4D Simplicial Quantum Gravity with $U(1)$
Gauge Field
\thanks{presented by S.Horata}}

\author{H.S.Egawa
        \address{Department of Physics, Tokai University, 
        Hiratsuka, Kanagawa 259-1292, Japan}
        $^{,}$
        \address{Theory Division, Institute of Particle 
        and Nuclear Studies, KEK, High Energy Accelerator 
        Research Organization, Tsukuba, Ibaraki 305-0801, Japan}
        ,
        S.Horata$^{\,\,{\rm b}\,\, , \,\, {\rm a}}$
        ,
        N.Tsuda$^{\,\, {\rm b}}$
        and 
        T.Yukawa 
        \address{Coordination Center for Research and Education, 
        The Graduate University for Advanced Studies, 
        Hayama, Miura, Kanagawa 240-0193, Japan}
        $^{\! , \,\,{\rm b}}$}
\begin{document}

\begin{abstract}
The phase transition of 4D simplicial quantum gravity coupled to $U(1)$
gauge fields is studied using Monte-Carlo simulations. 
The phase transition of the dynamical triangulation model with a vector field
($N_{V} = 1$) is smooth compared with pure gravity($N_{V}=0$).
The node susceptibility ($\chi$) is studied by the finite size scaling method.
At the critical point, the node distribution has a sharp peak in
 contrast to the double peak in pure gravity.
From the numerical results, we expect that 4D simplicial quantum gravity
with $U(1)$ vector fields has a higher order phase transition
than 1st order, which means the possibility to take the continuum limit
at the critical point.

\end{abstract}
\maketitle 
\section{Introduction} 
The phase structure of 4D pure simplicial quantum gravity has been
intensively investigated.
In 4D pure gravity, two distinct phases are known.
For small values of the bare gravitational coupling constant the 
system is in the so-called elongated phase, which has the characteristics 
of a branched polymer.
For large values of the bare gravitational coupling constant it is 
in the so-called crumpled phase.
Numerically, the phase transition between the two phases has been shown 
to be 1st order \cite{BBKP,BAKKER}.
As a result, it is difficult to construct a continuum theory. 
Our next step is to investigate an extended model of 4D quantum gravity.
Recently the phase structure of the extended models of 4D quantum
gravity has been studied numerically \cite{BILKE,PHASE-DIAGRAM}.
In a model with vector fields, the intermediate phase has
been observed between the crumpled phase and the elongated phase. 
We consider the possibility of a continuum limit at the critical point.
In order to investigate the phase transition, we measure the
finite size scaling of node susceptibility ($\chi$),
\begin{equation}
 \chi = (<N_0^2> - <N_0>^2)/N_4 ,
\end{equation}
as well as the scaling property of the mother boundary
in the case of $4$D simplicial quantum gravity coupled to one
gauge field ($N_{V}=1$).
%
\section{Phase Diagram with Gauge Fields}
We consider the partition function of simplicial gravity
coupled to $U(1)$ gauge fields.
The total action is $S = S_{EH} + S_{pl}$.
We use the Einstein-Hilbert action for gravity,
\begin{equation}
S_{EH}[\Lambda,G] = \displaystyle{\int} d^{4}x \sqrt{g}(\Lambda-\frac{1}{G}R), 
\end{equation}
where $\Lambda$ is the cosmological constant and $G$ is Newton's
constant.
We use discretized action for gravity,
\begin{equation}
 S_{EH}[\kappa_{2},\kappa_{4}] = \kappa_{4}N_{4}-\kappa_{2}N_{2},
\end{equation}
where $\kappa_{2} \sim 1/G$, $\kappa_{4}$ is related to $\Lambda$ and 
$N_{i}$ is the number of $i$-simplices.
We use the plaquette action for $U(1)$ gauge fields \cite{BILKE}, 
\begin{equation}
S_{pl}=\sum_{t_{ijk}}o(t_{ijk})[A(l_{ij})+A(l_{jk})+A(l_{ki})]^{2},
\end{equation}
where $l_{ij}$ denotes a link between vertices $i$ and $j$, $t_{ijk}$ denotes 
a triangle with vertices $i$, $j$ and $k$, $A(l_{ij})$ denotes the $U(1)$ 
gauge field on a link $l_{ij}$, and $o(t_{ijk})$ denotes the number of
4-simplices sharing triangle $t_{ijk}$.
We consider that a partition function of gravity with $N_V$ copies of
the $U(1)$ gauge fields is
\begin{eqnarray}
 Z(\kappa_{2},\kappa_{4},N_V) &=& \sum_{N_4} e^{-\kappa_4 N_4} \sum_{t(2D) \in T(4D)} e^{\kappa_2 N_2} \nonumber \\
& & \prod_{N_V}
  \int \prod_{l \in t(2D)} dA(l) e^{-S_{pl}}.
\end{eqnarray}
We sum over all 4D simplicial triangulation, 
$T(4D)$, in order to carry out a path integral over the metric. 
Here, we fix the topology with $S^{4}$.
Numerically, in the case of adding vector fields, three phases have
been found \cite{BILKE,PHASE-DIAGRAM}.
A schematic phase diagram has been shown in ref.\cite{PHASE-DIAGRAM}.
%
%
An intermediate region is called the smooth phase between these 
two transition points,\footnote[1]{The usual phase transition point
($\kappa_{2}^{c}$) and the obscure phase transition point
($\kappa_{2}^{o}$) are defined in ref.\cite{PHASE-DIAGRAM}}
$\kappa_{2}^{c}$ and $\kappa_{2}^{o}$.
%
%
We expect that the phase transition at $\kappa_{2}^{c}$ is
continuous and leads to continuum limit of 4D quantum gravity.
Now, let us notice the transition at $\kappa_{2}^{c}$ in the
case of $N_V=1$.
%
\section{Numerical Analysis of Phase Transition ($N_{V}=1$)}
%
In this section we report on two numerical observations: the node
susceptibility ($\chi$) and the histogram of $N_0$.
In Fig.1 we plot the node susceptibility ($\chi$) as a function of
$\kappa_2$ with volume $N_{4}=16K,24K$ and $32K$, respectively.
The node susceptibility ($\chi$) has a peak value at the critical point
($\kappa_2^c$).
We measure the peak value in each size.
As a finite size scaling, the peak value ($\chi_{max}$) and the width of 
peak ($\delta \kappa_2$) grow as $N_4$ in power.
The susceptibility exponents, $\Delta$ and $\Gamma$, are defined by
\cite{BAKKER}
\begin{equation}
 \chi_{max} \propto N_{4}^{\Delta}
 \quad (\delta \kappa_2 \propto N_4^{-\Gamma}).
\end{equation}
From the numerical result (Fig.1), we obtain the susceptibility
exponent $\Delta = 0.4(1)$, ($\Gamma \sim 0.5(3)$).
These values are apparently smaller than 1.
In Fig.2 we show the histogram of $N_0$ at the critical point
($\kappa_2^{c} = 1.37147(1)$) with the size of $N_{4} =
32K$.
In pure gravity, a double peak structure has been
found\cite{BBKP,BAKKER}. The fact shows that the phase transition is 1st
order.
However, in the case of $N_V=1$, the double peak structure disappears.
We then consider that the phase transition between the crumpled phase
and the smooth phase may be continuous, not 1st order.
%
\section{Scaling Property of 4D Simplicial Quantum Gravity}
%
\begin{figure}[t]
\centerline{\psfig{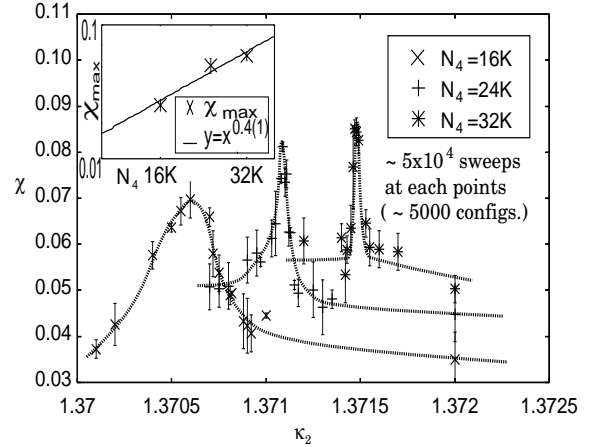}} 
\vspace{-10mm}
\caption
{
Node susceptibility ($\chi$) plotted versus the 
coupling constant ($\kappa_{2}$) and $\chi_{max}$ plotted versus $N_4$
with log-log scales for $N_{V}=1$.
}
\label{fig:fss}
\vspace{-0.7cm}
\end{figure}
In this section we discuss the scaling structure of 4D DT manifolds,
focusing on the scaling structure of boundaries in 4D Euclidean space-time
using the concept of geodesic distances. 
In order to discuss the universality of the scaling relations, we 
assume that the boundary volume distribution ($\rho(x,D)$) is a function
of a scaling variable, $x=V/D^{\alpha}$, with the scaling parameter
$\alpha$ in the analogy of 2D quantum gravity.
Here, $V$ denotes the volume of the boundary and $D$ is the geodesic distance.
The expectation value of the boundary three-dimensional volume appearing
at distance $D$ has been introduced in ref.\cite{EHITY},
\begin{equation}
 <V^{(3)}>=\frac{1}{N} \int_{v_0}^{\infty} dV \, V \rho(x=V/D^\alpha,D),
\end{equation}
where $v_0$ denotes UV cut-off of the boundary volume and $N$ is
the normalization factor.
%
\begin{figure}
\centerline{\psfig{file=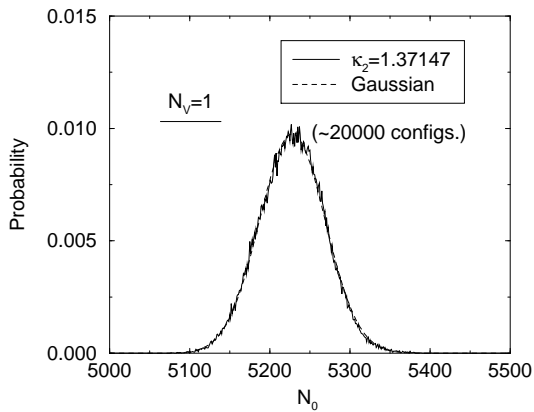,height=5.5cm,width=7cm}} 
\vspace{-10mm}
\caption
{
Histogram of node($N_0$) at the critical point ($\kappa_2=1.37147$) with the
 size of $N_4=32K$.
}
\label{fig:node}
\vspace{-0.7cm}
\end{figure}
If the boundary volume has a scaling property with the universal
distribution $\rho(x,D)$ and $v_0 \rightarrow 0$,
\begin{equation}
 <V^{(3)}> \sim D^{\alpha}.
\end{equation}
Then, we obtain a finite fractal dimension, $d_f = \alpha + 1$, with the
fractal dimension $d_f$.
We measure the volume of the mother boundary as a function of $D$.
The mother boundary is defined by the boundary having the largest tip
volume.
In Fig.3 we plot the mother boundary volume, $<V^{(3)}>$, at the critical
point with a size of $N_4 = 32K$.
As a result, the mother boundary volume shows scaling and
we obtain the  scaling parameter $\alpha = 3.7(5)$.
Then, we can estimate the fractal dimension ($d_f= 4.7(5)$).
On the other hand, we measure the Hausdorff dimension, $d_H = 4.6(2)$.
Both results are consistent. 
%
%
Thus, we expect that the boundary volume distribution has a scaling
property in the sense of the manifold at different distances from a
given $4$-simplex and looks exactly the same after a proper rescaling of
the boundary volume.
%
%
\section{Summary and Discussions}
%
\begin{figure}[t]
\centerline{\psfig{file=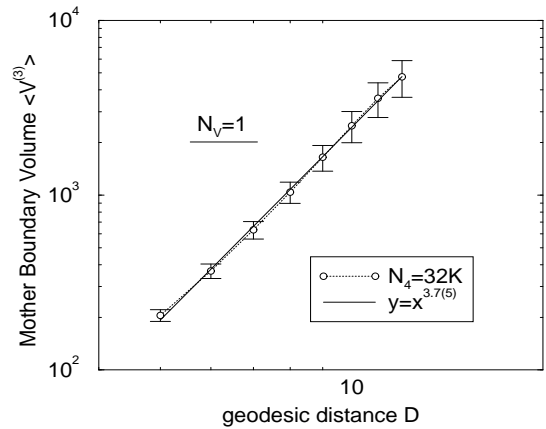,height=5.7cm,width=7cm}} 
\vspace{-10mm}
\caption
{
Power fit to mother boundary three-dimensional volume $<V^{(3)}>$ at the 
 critical point ($\kappa_2=1.37147$) with the size of $N_4=32K$ with
 log-log scales.
}
\label{fig:mother}
\vspace{-0.7cm}
\end{figure}
Let us summarize the main points made in the previous sections.
%
%
For the phase transition,
we show the finite size scaling at the critical point and the histogram
of node. 
From the numerical results, the phase transition is smooth in contrast
to pure gravity.
Also we show the scaling property of the mother boundary, where the scaling
parameter is consistent with the fractal dimension.
We expect that the boundaries have a fractal structure and the universality
of the scaling relations.
From a modification of the Balls-in-Boxes model\cite{BB}, we expect
that the simplicial quantum gravity coupled to matter fields will have the
possibility of a continuous phase transition in $N_V \ge 1$.
We expect that the existence of genuine 4D quantum gravity at the critical
point, $\kappa_{2}^{c}$, with the vector fields.

\end{document}